\documentclass[journal]{IEEEtran}
\newcommand{\RNum}[1]{\uppercase\expandafter{\romannumeral #1\relax}}
\usepackage{cite}
\usepackage{amsmath,amssymb,amsfonts,bm}
\newtheorem{assumption}{Assumption}
\usepackage{algorithmic}
\usepackage{dsfont}
\usepackage{graphicx}
\usepackage{float}
\usepackage[caption=false]{subfig}
\usepackage{epstopdf}
\usepackage{textcomp}
\usepackage{xcolor}
\usepackage{multirow}
\usepackage{algorithm}
\usepackage{algorithmic}
\newtheorem{theorem}{Theorem}

\usepackage{tabularx}
\usepackage{hyperref}
\hypersetup{colorlinks=true}
\hypersetup{linkcolor=blue}
\hypersetup{filecolor=blue}  
\hypersetup{citecolor=blue} 
\hypersetup{urlcolor=black}

\begin{document}
\newenvironment{shrinkeq}[1]
{ \bgroup
\addtolength\abovedisplayshortskip{#1}
\addtolength\abovedisplayskip{#1}
\addtolength\belowdisplayshortskip{#1}
\addtolength\belowdisplayskip{#1}}
{\egroup\ignorespacesafterend}
	
\title{Federated Contrastive Learning for Personalized Semantic Communication}

\author{Yining~Wang,
		Wanli~Ni,
		Wenqiang~Yi,~\IEEEmembership{Member,~IEEE,}
        Xiaodong~Xu,~\IEEEmembership{Senior~Member,~IEEE,}
		Ping~Zhang,~\IEEEmembership{Fellow,~IEEE,}
        and~Arumugam~Nallanathan,~\IEEEmembership{Fellow,~IEEE}% <-this % stops a space
        \vspace{-6 mm}
		\thanks{The work presented in this paper is funded by the National Key R\&D Program of China No. 2020YFB1806905, the National Natural Science Foundation of China No. 62201079, the Beijing Natural Science Foundation No. L232051 and the Major Key Project of PCL Department of Broadband Communication. \textit{(Corresponding author: Xiaodong Xu.)}}
		\thanks{Yining Wang is with the State Key Laboratory of Networking and Switching Technology, Beijing University of Posts and Telecommunications, Beijing, 100876, China (e-mail: \href{mailto:joanna_wyn@bupt.edu.cn}{joanna\_wyn@bupt.edu.cn}).}% <-this % stops a space
        \thanks{Wanli Ni is with Department of Electronic Engineering, Tsinghua University, Beijing, 100084, China (e-mail: \href{mailto:niwanli@tsinghua.edu.cn}{niwanli@tsinghua.edu.cn}).}% <-this % stops a space
        \thanks{Wenqiang Yi is with the School of Computer Science and Electronic Engineering, University of Essex, Colchester CO4 3SQ, U.K. (e-mail: \href{mailto:wy23627@essex.ac.uk}{wy23627@essex.ac.uk}).}% <-this % stops a space
		\thanks{Xiaodong Xu and Ping Zhang are with the State Key Laboratory of Networking and Switching Technology, Beijing University of Posts and Telecommunications, Beijing 100876, China, and also with Peng Cheng Laboratory, Shenzhen 518066, China (e-mail: \href{mailto:xuxiaodong@bupt.edu.cn}{xuxiaodong@bupt.edu.cn}; \href{mailto:pzhang@bupt.edu.cn}{pzhang@bupt.edu.cn}). }% <-this % stops a space
        \thanks{Arumugam Nallanathan is with the School of Electronic Engineering and Computer Science, Queen Mary University of London, London E1 4NS, U.K. (e-mail: \href{mailto:a.nallanathan@qmul.ac.uk}{a.nallanathan@qmul.ac.uk}).}% <-this % stops a space
		}
\maketitle
\begin{abstract}
In this letter, we design a federated contrastive learning (FedCL) framework aimed at supporting personalized semantic communication. Our FedCL enables collaborative training of local semantic encoders across multiple clients and a global semantic decoder owned by the base station. This framework supports heterogeneous semantic encoders since it does not require client-side model aggregation. Furthermore, to tackle the semantic imbalance issue arising from heterogeneous datasets across distributed clients, we employ contrastive learning to train a semantic centroid generator (SCG). This generator obtains representative global semantic centroids that exhibit intra-semantic compactness and inter-semantic separability. Consequently, it provides superior supervision for learning discriminative local semantic features. Additionally, we conduct theoretical analysis to quantify the convergence performance of FedCL. Simulation results verify the superiority of the proposed FedCL framework compared to other distributed learning benchmarks in terms of task performance and robustness under different numbers of clients and channel conditions, especially in low signal-to-noise ratio and highly heterogeneous data scenarios.
\end{abstract}
	
	
% \fi
\IEEEpeerreviewmaketitle

\section{Introduction}
\IEEEPARstart{T}{ASK}-oriented semantic communication (SemCom) systems mainly employ sophisticated deep neural network (DNN) models or optimize wireless resource allocation to balance communication efficiency with target performance. However, few of them addressed the training approach of DNN-based semantic models, while the effectiveness of task-oriented SemCom relies heavily on semantic models deployed on each transceiver, which requires continuous update along with the changing channel environment and datasets \cite{lu2023semantics}.

Since semantic model learning requires a huge quantity of training samples from dispersed users, most existing works exploited federated learning (FL) approaches. \cite{xing2023multi} proposed a FL-based semantic learning system with dynamic model aggregation. However, it relied on local training and central server aggregation, leading to high parameter transmission and underutilization of the server's computing power. In \cite{nguyen2023efficient}, a FL framework for semantic reconstruction reduced communication costs through partial client model aggregation, but still conducted training locally, neglecting server's potential. Wei \textit{et al.} in \cite{wei2023federated} introduced a client-server collaborative FL framework for knowledge graph generation, yet still requiring model uploading and client-side transmitter aggregation.

\begin{figure*}[ht]
    \centering
		\subfloat[]{
            \label{model}
			\includegraphics[width=0.67\textwidth]{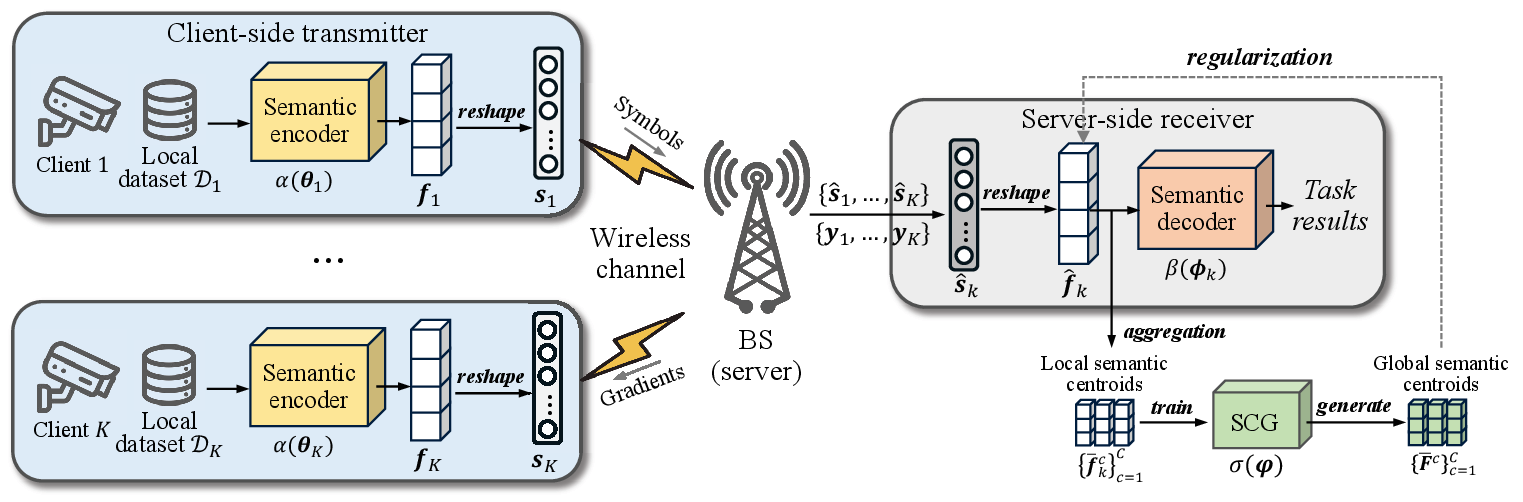}
		}
        \subfloat[]{
            \label{FedContrast}
			\includegraphics[width=0.3\textwidth]{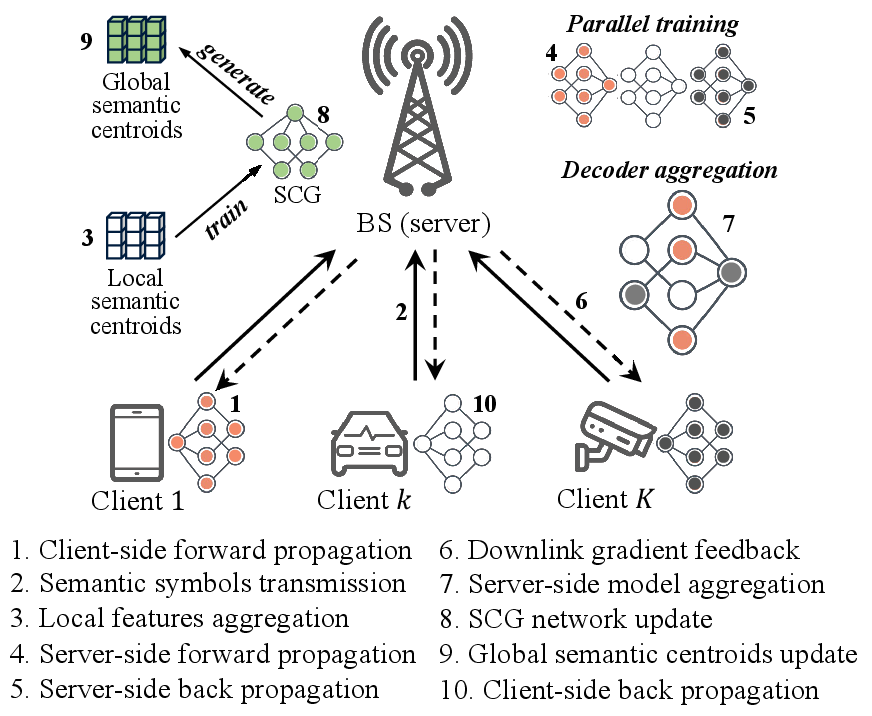}
		}
		\caption{(a) Architecture of the proposed FedCL for multi-user semantic learning; (b) Workflow of the proposed FedCL framework.}
  \label{system}
\end{figure*}
The previous studies assumed uniformity in user semantic models, limiting their use to homogeneous settings. However, in practice, client-side semantic transmitters should accommodate personalized encoders, which can adapt to diverse data distributions and varying model structures due to local devices' different computation and storage capabilities. Moreover, existing research ignored the non-independent and identically distribution (non-IID)  data among users. This inconsistency in feature spaces across clients degrades the performance of traditional FL \cite{ye2023fedfm,t2020personalized}.
By grouping intra-class samples as positives and distinguishing inter-class samples as negatives, contrastive learning \cite{khosla2020supervised} fosters the learning of discriminative features that aid in identifying semantics, even in scenarios with unbalanced data distributions \cite{jing2023exploring}. Applying this principle, contrastive loss can guide the training of semantic models by generalizing knowledge from similar samples while minimizing interference from semantically inconsistent ones \cite{tan2022federated}.

In this work, we propose a federated contrastive learning (FedCL) framework for task-oriented SemCom, where personalized semantic encoders and a global semantic decoder are trained collaboratively between the clients and the base station (BS). The main contributions of this work are summarized as follows:

\begin{itemize}
\item[$\bullet$] We design a novel FedCL framework for collaborative training of personalized semantic encoders on multiple clients and a global semantic decoder on the BS. Instead of exchanging model parameters or raw data, our approach exchanges features and back-propagation gradients, which not only preserves user privacy but also eliminates client-side model aggregation.
\item[$\bullet$] To overcome performance degradation from inconsistent semantic distributions in heterogeneous multi-user datasets, we introduce a semantic centroid generator (SCG) at the server. This network leverages contrastive learning to generate global semantic centroids, which are updated in each round to provide a unified semantic space for supervised local semantic feature learning. This approach transforms noisy features from heterogeneous data distributions into regularized features with intra-semantic compactness and inter-semantic separability, thereby enhancing robustness against channel noise.
\item[$\bullet$] We theoretically analyze the convergence performance of FedCL under the non-convex loss function setting, which provides a convergence guarantee to the proposed framework. Simulation results demonstrate that FedCL surpasses benchmark approaches in task performance, particularly in scenarios with low signal-to-noise ratio (SNR) and significant data heterogeneity.
\end{itemize}

\section{System Model}
\label{system model}
We consider a wireless network comprising one BS with an edge server and a set of devices $\mathcal{K}=\{1,2,...,K\}$. The clients and the BS learn collaboratively to obtain personalized semantic encoders for feature extraction as well as channel encoding on each client, and a global semantic decoder for channel decoding as well as performing downstream tasks among semantic concepts $\mathcal{C}=\{1,2,...,C\}$. In this section, we propose a FedCL framework for personalized semantic communication, which facilitates the training of heterogeneous semantic models among distributed clients. 
\vspace{-3 mm}
\subsection{FedCL Framework}
As depicted in Fig. \ref{system}\subref{model}, a semantic encoder is deployed on client $k$ to facilitate feature extraction from raw data while considering the impact of wireless channel. The local dataset $\mathcal{D}_k$ with $D_k$ data samples owned by client $k$ is denoted as $\mathcal{D}_k=\{(\boldsymbol{x}_{k,i},y_{k,i})|i=1,2,...,D_k\}$, where $\boldsymbol{x}_{k,i}$ represents the source input and $y_{k,i}$ is the corresponding label indicating the semantic concept of $\boldsymbol{x}_{k,i}$. Thus, the entire dataset $\mathcal{D}$ of all clients is denoted by $\mathcal{D}=\cup_{k=1}^K\mathcal{D}_k$ with $D=\sum_{k=1}^K{D_k}$ data samples. Note that the subscript $i$ is omitted when sample-wise formulation is not required.
We consider a training process for FedCL with $T$ communication rounds, as shown in Fig. \ref{system}\subref{FedContrast}. Specifically, in the $t$-th round, the FedCL process consists of the following stages.

First, each client $k$ performs the forward propagation in parallel on its personalized semantic encoder and extracts feature $\boldsymbol{f}_{k}$ from input data $\boldsymbol{x}_{k}$, which is denoted as
\begin{align}
\label{feature}
\boldsymbol{f}_{k}=\alpha(\boldsymbol{x}_{k};\boldsymbol{\theta}_k^{(t)}), \forall k \in \mathcal{K},
\end{align}
where $\boldsymbol{\theta}_k^{(t)}$ is the semantic encoder parameter set of client $k$ in the $t$-th round. After the client-side forward propagation is completed, the encoded feature $\boldsymbol{f}_{k}$ is reshaped into semantic symbols $\boldsymbol{s}_{k}$ and transmitted to the BS over the wireless channel along with the label $\boldsymbol{y}_{k}$. We assume the proposed learning system is operated with orthogonal frequency division multiplexing (OFDM), where the channel is divided into orthogonal subcarriers according to the number of participating users. Thus, it ensures the access of multiple local devices without interference.
Then, the feature received at the BS can be expressed as
\begin{align}
\hat{\boldsymbol{s}}_{k}=h_k\boldsymbol{s}_k+\boldsymbol{n}_k, \forall k \in \mathcal{K},
\end{align}
where $h_k$ denotes the channel coefficient and $\boldsymbol{n}_k$ is the IID channel noise vector, which follows symmetric complex Gaussian distribution $\mathcal{CN}(0,\delta^2\boldsymbol{I})$ with zero mean and variance $\delta^2$ \cite{10388062}. Note that since the data volume of the label $\boldsymbol{y}_{k}$ is small, it can be transmitted accurately to the server.

Subsequently, the noised semantic symbols received from all participating clients are reshaped into noised semantic feature $\hat{\boldsymbol{f}}_k$, which is utilized as the input for forward propagation of the semantic decoder hosted on the BS. In parallel, the BS trains a separate instance $\beta(\boldsymbol{\phi}_k)$ of the semantic decoder for each user, which outputs the task results as
\begin{align}
\label{task result}
r_{k}=\beta(\hat{\boldsymbol{f}}_{k};\boldsymbol{\phi}_k^{(t)}), \forall k \in \mathcal{K},
\end{align}
where $\boldsymbol{\phi}_k^{(t)}$ is the parameter set of client $k$'s semantic decoder in the $t$-th round. The parallel training method is adopted at the server side, where the semantic decoder parameters of each client can be iteratively updated using stochastic gradient descent (SGD). In the $t$-th communication round, the updated parameters of the $k$-th semantic decoder can be obtained as
\begin{align}
\label{decoder update}
\boldsymbol{\phi}_k^{(t+1)}=\boldsymbol{\phi}_k^{(t)}-\eta \boldsymbol{g}_k^{(t)}, \forall k \in \mathcal{K},
\end{align}
where $\eta$ is the learning rate of the semantic decoder network and $\boldsymbol{g}_k^{(t)}$ is the obtained gradient in the $t$-th communication round. At the end of each round, model aggregation is performed at the BS to obtain an updated global semantic decoder $\beta(\boldsymbol{\phi})$. The decoder parameters in $t+1$ are aggregated as,
\begin{align}
\label{decoder aggregation}
\boldsymbol{\phi}^{(t+1)}=\sum_{k=1}^K \frac{D_k}{D}\boldsymbol{\phi}_k^{(t+1)},
\end{align}

When the back propagation process reaches the first layer of the semantic decoder, the BS sends the updated gradients $\boldsymbol{g}_k^{(t)}$ to all participating clients over the wireless channel to guide the back propagation of the personalized semantic encoders. At the client's side, after receiving the noised gradients $\check{\boldsymbol{g}}_k^{(t)}$ which is corrupted by the downlink transmissions between the BS and the devices, each client $k$ performs back propagation on its local model and updates the parameters using the gradient descent method, i.e., 
\begin{align}
\label{encoder update}
\boldsymbol{\theta}_k^{(t+1)}=\boldsymbol{\theta}_k^{(t)}-\eta_k\check{\boldsymbol{g}}_k^{(t)}, \forall k \in \mathcal{K},
\end{align}
where $\eta_k$ denotes the learning rate of client $k$. The local learning rate can vary among different clients with heterogeneous computing power and latency requirements. After completing the back propagation process on the client side, the updated semantic encoder of round $t+1$ is obtained as $\alpha(\boldsymbol{\theta}_k^{(t+1)})$.
\subsection{Contrastive Learning-Based SCG}
In real-world scenarios, clients deployed across various locations encounter diverse environments, resulting in semantic heterogeneity among local datasets. This discrepancy, known as non-IID distribution, arises as each client's semantic distribution is inconsistent with the server. To address this statistical heterogeneity issue, we employ a contrastive learning method to align the inconsistent semantic space across clients into unified global semantic distribution by training a semantic centroid generator (SCG).

Each client has its local semantic centroid $\overline{\boldsymbol{f}}_k^c$ for each semantic concept $c$, which is aggregated on the server as
\begin{align}
\label{local aggregation}
    \overline{\boldsymbol{f}}_k^c=\frac{1}{D_{k,c}}\sum_{i\in\mathcal{D}_{k,c}}\hat{\boldsymbol{f}}_{k,i}, \forall k\in\mathcal{K}, c\in\mathcal{C},
\end{align}
where $\mathcal{D}_{k,c}$ denotes the set of samples belong to $c$-th category on client $k$ with $D_{k,c}$ data samples. 

However, due to the statistical and model heterogeneity of personalized semantic encoders, the aggregated semantic centroids of different clients are much diverse even if they are with the same semantic concept. Therefore, dislike other FL frameworks with centroid regularization that achieve the global centroids by simply aggregating the local centroids \cite{chen2023knowledge,tan2022fedproto}, we design the SCG to generate trainable global semantic centroids $\overline{\boldsymbol{F}}=\{\overline{\boldsymbol{F}}^c\}_{c=1}^C$ via contrastive learning. The proposed SCG is constructed by two fully-connected layers with ReLU activation in the middle, and such structure is proven useful in improving the quality of representations \cite{chen2020simple}.

Specifically, we first randomly initialize each global semantic centroid vector. Then the SCG model $\sigma(\cdot)$ parameterized by $\boldsymbol{\varphi}$ is updated in each round to generate better global semantic centroids using the following objective
\begin{align}
\label{min scg}
\min_{\boldsymbol{\varphi}}\sum_{c=1}^C\mathcal{L}_F^c,
\end{align}
\vspace{-3 mm}
\begin{align}
\label{scg loss}
\mathcal{L}_F^c\!=\!\sum_{k=1}^K\underbrace{\log\!\big(\!{\sum_{n\in \mathcal{N}_c}\!\exp\!{(\overline{\boldsymbol{f}}_k^c\!\cdot\! {\overline{\boldsymbol{F}}^n})}}\big)}_{\textit{loss for negatives}}\!-\!\underbrace{\log\big({\exp{(\overline{\boldsymbol{f}}_k^c\!\cdot\!{\overline{\boldsymbol{F}}^c})}}\big)}_{\textit{loss for positives}},
\end{align}
where $\mathcal{N}_c$ denotes the set of semantic concepts other than $c$. Therefore, by maximizing the similarity between each local semantic centroid $\overline{\boldsymbol{f}}_k^c$ and the global semantic centroid $\overline{\boldsymbol{F}}^c$ of its ground-truth semantic concept $c$ (positives), while simultaneously minimizing the similarity between $\overline{\boldsymbol{f}}_k^c$ and the global semantic centroids of other irrelevant semantic concepts (negatives), the SCG can generate representative global semantic centroids that preserve the semantic information while maintaining certain distance from centroids with different semantics.

Deriving the SCG model on the server, global semantic centroids are generated with better inter-semantic separability and intra-semantic compactness, which are exploited for the regularization of noised local semantic features. Thus, each semantic model is guided by a regularized loss function as
\begin{align}
\label{lossk}
    \mathcal{L}_k=\frac{1}{D_k}\sum_{c=1}^{C}\sum_{i\in\mathcal{D}_{k,c}}\mathcal{L}_{T}(r_{k,i},y_{k,i})+\lambda\|\hat{\boldsymbol{f}}_{k,i}-\overline{\boldsymbol{F}}^c\|_2^2,
\end{align}
where $\mathcal{L}_T$ denotes the task loss and $\lambda$ is the regularization coefficient. Thus, the optimization goal of the entire FedCL framework is expressed as
\begin{align}
    \min_{\{\boldsymbol{\theta}_k,\boldsymbol{\phi}_k\}_{k=1}^K,\boldsymbol{\varphi}}\frac{1}{K}\sum_{k=1}^K\mathcal{L}_k.
\end{align}
Under the L2 supervision of SCG-based global semantic centroids, all noised semantic features from different clients with heterogeneous data distributions and channel conditions are restricted in a consistent global semantic space, thereby integrating the personalized semantic features and preserving the shared semantics in a compact form.

\begin{algorithm}[t]
\normalsize
\caption{Training process of the FedCL framework}
\begin{algorithmic}[1]
\STATE \textbf{Input:} Dataset $\mathcal{D}_k$ of each client $k$, noised channel generated from a fixed distribution. 
\STATE \textbf{Initialize:} Client-side semantic encoder parameters $\boldsymbol{\theta}_k^{(0)}$, server-side semantic decoder parameters $\boldsymbol{\phi}_k^{(0)}$, global semantic centroids $\overline{\boldsymbol{F}}=\{\overline{\boldsymbol{F}}^c\}_{c=1}^C$.
\WHILE{communication round $t=0$ to $200$}
\FOR{$k=1$ to $K$}
\STATE Extract $\boldsymbol{f}_k$ using semantic encoder $\alpha(\boldsymbol{\theta}_k^{(t)})$ by \eqref{feature}.
\STATE Reshape $\boldsymbol{f}_k$ as $\boldsymbol{s}_k$ and transmit over the channel.
\STATE Receive and reshape the noised symbols $\hat{\boldsymbol{s}}_k$ at the server and obtain $\hat{\boldsymbol{f}}_k$.
\STATE Output task result $r_{k}$ using \eqref{task result} at the BS.
\STATE Aggregate local semantic centroids $ \{\overline{\boldsymbol{f}}_k^c\}_{c=1}^C$ by \eqref{local aggregation}.
\STATE Calculate $\mathcal{L}_k$ using \eqref{lossk} at the BS server.
\STATE Update semantic decoder by \eqref{decoder update} and obtain $\boldsymbol{\phi}_k^{(t+1)}$.
\STATE Transmit gradients $\boldsymbol{g}_k^{(t)}$ back to client $k$ over downlink wireless channel and obtain $\check{\boldsymbol{g}}_k^{(t)}$.
\STATE Update semantic encoder by \eqref{encoder update} and obtain $\boldsymbol{\theta}_k^{(t+1)}$.
\ENDFOR
\STATE Update the SCG on server using \eqref{min scg} and update global semantic centroids $\overline{\boldsymbol{F}}$.
\STATE Aggregate server-side semantic decoder by \eqref{decoder aggregation}.
\ENDWHILE
\STATE \textbf{Output:} Converged semantic encoder and decoder model.
\end{algorithmic}
\label{training process}
\end{algorithm}
Note that the SCG is deployed on the server but independent of the semantic decoder, which is only used to generate global semantic centroids as supervision during semantic model training. There is no other parameter interactions between the SCG and the semantic decoder.
The entire training process of the proposed FedCL framework is described in Algorithm \ref{training process}.
\vspace{-5mm}

\subsection{Convergence Analysis}
Denoting the entire semantic model parameters of client $k$ as $\boldsymbol{w}_k=\{\boldsymbol{\theta}_k,\boldsymbol{\phi}_k\}$, we analyze the convergence performance of FedCL by introducing the following assumptions:
\begin{assumption}
\label{a1}
(Lipschitz smooth). Each loss function is $L_1$-Lipschitz smooth, and the gradient of each loss function is $L_1$-Lipschitz continuous. Since this assumption is valid for arbitrary client, we omit the footnote $k$,
\begin{align}
\begin{split}
	\|\nabla\mathcal{L}_{t_1}-\nabla\mathcal{L}_{t_2}\|_2\leq L_1\|\boldsymbol{w}_{t_1}-\boldsymbol{w}_{t_2}\|_2,\forall t_1, t_2>0.
\end{split}
\end{align}
This also implies the following quadratic bound,
\begin{align}
	\mathcal{L}_{t_1}\!-\!\mathcal{L}_{t_2}\!\leq\!\langle\nabla\mathcal{L}_{t_2},(\boldsymbol{w}_{t_1}\!-\!\boldsymbol{w}_{t_2})\rangle\!+\!\frac{L_1}{2}\|\boldsymbol{w}_{t_1}\!-\!\boldsymbol{w}_{t_2}\|_2^2.
\end{align}
\end{assumption}
\begin{assumption}
\label{a2}
(Unbiased gradient and bounded variance). The stochastic gradient $g_t=\nabla\mathcal{L}(\boldsymbol{w}_t,\xi_t)$ is an unbiased estimator of the local gradient for each client $k$. Its expectation is formulated as
\begin{align}
	\mathbb{E}_{\xi_k\sim\mathcal{D}_k}[\boldsymbol{g}_{k,t}]=\nabla\mathcal{L}_k(\boldsymbol{w}_{k,t})=\nabla\mathcal{L}_t, \;\;
	\forall k\in\mathcal{K},
\end{align}
and its variance is bounded by $\rho^2$:
\begin{align}
	\mathbb{E}[\|\boldsymbol{g}_{k,t}-\nabla\mathcal{L}(\boldsymbol{w}_{k,t})\|_2^2]\leq\rho^2, \;\;
	\forall k\in\mathcal{K}.
\end{align}
\end{assumption}
\begin{assumption}
\label{a3}
(Lipschitz continuity). The SCG network $\sigma(\boldsymbol{\varphi})$ is $L_2$-Lipschitz continuous, that is,
\begin{align}
	\|\sigma(\boldsymbol{\varphi_{t_1}})-\sigma(\boldsymbol{\varphi_{t_2}})\| \leq L_2\|\boldsymbol{\varphi_{t_1}}-\boldsymbol{\varphi_{t_2}}\|_2, \;\;
	\forall t_1,t_2>0.
\end{align}
\end{assumption}
\begin{assumption}
\label{a4}
(Bounded expectation of Euclidean norm of stochastic gradients). The expectation of the stochastic gradient of the SCG network is bounded by $G$:
\begin{align}
	\mathbb{E}[\|\boldsymbol{g}_{t}^\prime\|_2]\leq G.
\end{align}
\end{assumption}

Then we derive the expected one-round decrease in Theorem \ref{theorem1}. We denote $\{1/2,1,2,...,E\}$ as the local iteration of semantic model parameters $\boldsymbol{w}_k$, $\{1/2,1,2,...,E^\prime\}$ as the local iteration of SCG network parameters $\boldsymbol{\varphi}$, and $t$ as the global communication round. Additionally, $tE$ denotes the steps before global semantic centroid generating, and $tE+1/2$ represents the step between global semantic centroid generating and the first iteration of round $t$.
\begin{theorem}
\label{theorem1}
(One-round deviation bound). Let Assumptions \ref{a1} to \ref{a4} hold. For arbitrary client after each round, it satisfies,
\begin{align}
\label{eqtheorem1}
	\mathbb{E}[\mathcal{L}_{(t+1)E+1/2}]&\!\leq\!\mathcal{L}_{tE+1/2}\!-\!(\eta\!-\!\frac{L_1\eta^2}{2})\!
	\sum_{e=1/2}^{E-1}\!\|\nabla\mathcal{L}_{tE+e}\|_2^2 \nonumber \\
 &+\!\frac{L_1E\eta^2}{2}\rho^2\!+\!\lambda L_2\eta E^\prime G.
\end{align}
\end{theorem}

Theorem \ref{theorem1} exhibits a deviation bound of loss function for arbitrary client after each communication round. As observed in \eqref{eqtheorem1}, the second term on the right side is always negative. By tuning the value of $\lambda$ and $\eta$ according to the non-IID degree (for example, $\lambda=0$ is equivalent to vanilla FL with IID data distribution), a specific one-round decrease is obtained to guarantee the monotonic decrease in all communication rounds, thereby ensuring the convergence of FedCL.
\footnote{Complete proof is available at https://github.com/wangyining98/FedCL}

\section{Simulation Results}
\begin{figure}[t]
    \centering
		\subfloat[Learning performance vs. \textit{K}]{
            \label{test_acc}
			\includegraphics[width=0.24\textwidth]{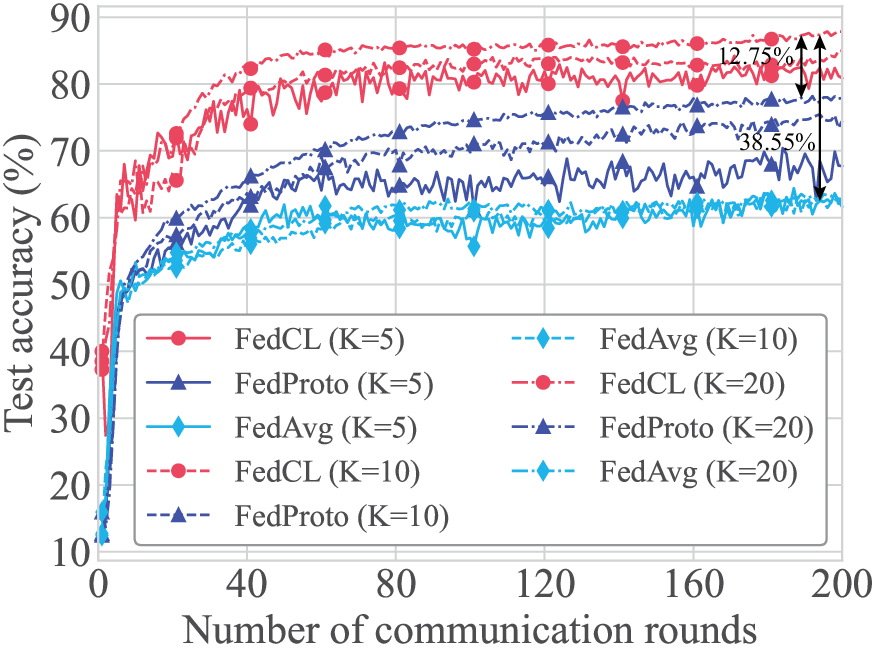}
		}
        \subfloat[Task performance vs. SNR]{
            \label{snr_acc}
			\includegraphics[width=0.24\textwidth]{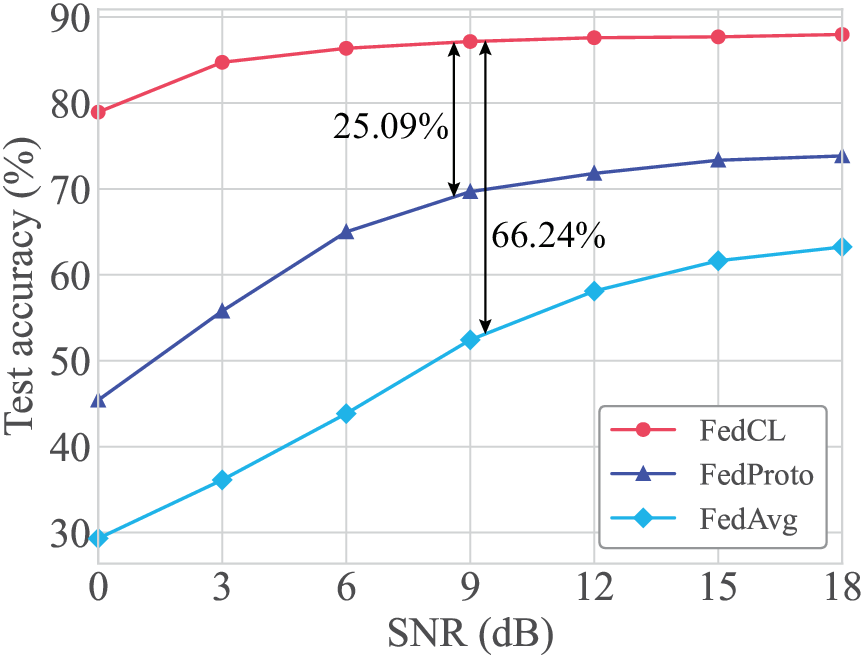}
		}
		\caption{(a) Learning performance of different schemes; (b)Task performance under different SNR with 20 clients.}
  \label{performance1}
\end{figure}
\begin{figure}[t]
    \centering
		\subfloat[Learning performance vs. \textit{m}]{
            \label{test_acc_m}
			\includegraphics[width=0.24\textwidth]{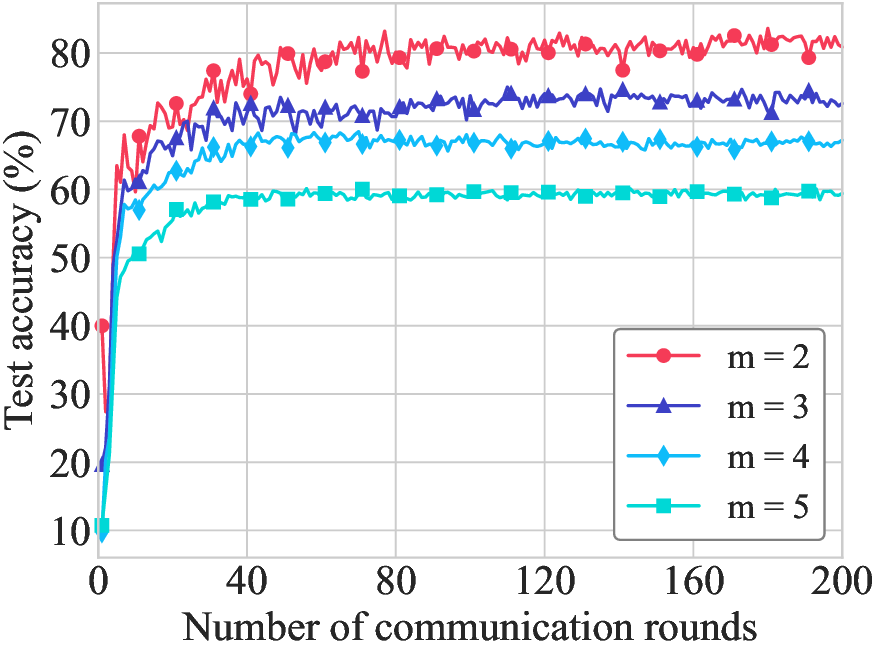}
		}
        \subfloat[Task performance vs. \textit{m}]{
            \label{m_acc}
			\includegraphics[width=0.24\textwidth]{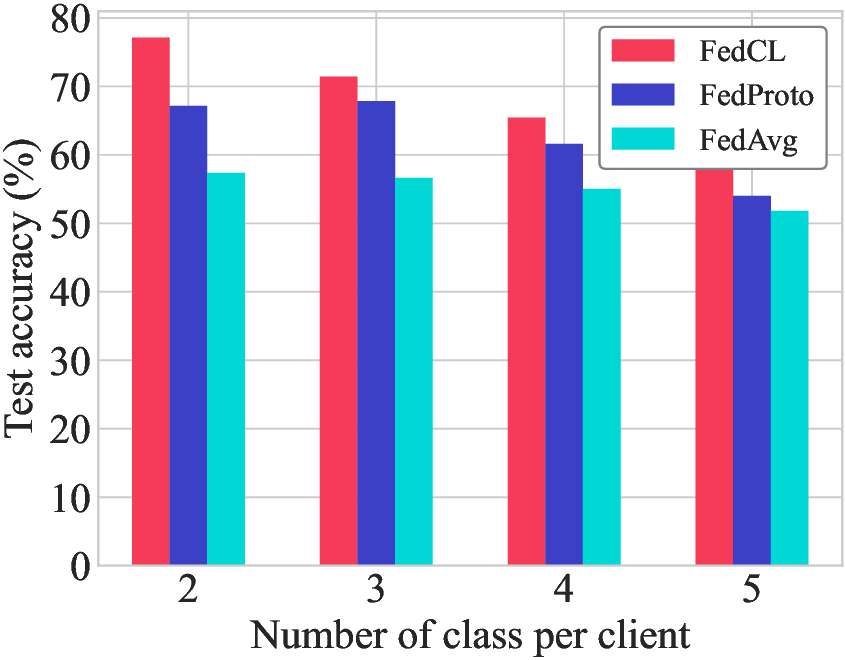}
		}
		\caption{(a) Learning performance under varying data semantic heterogeneity $m$ with 5 clients; (b) Task performance of different schemes with varying $m$.}
  \label{performance2}
\end{figure}
To verify the effectiveness of the FedCL framework, we compare the proposed scheme with conventional federated learning frameworks FedAvg \cite{mcmahan2017communication} and FedProto \cite{tan2022fedproto} on CIFAR-10 dataset. The semantic encoder is designed as DCGAN-like encoders \cite{radford2015unsupervised} with output dimension of 64. The SCG is constructed by two fully connected layers with 64 outputs, and the semantic decoder is designed as a fully-connected image classification network with 10-unit outputs. The learning rates, i.e., $\eta_k$ and $\eta$ are set to $0.001$ for both clients and server.

We divide the client dataset as $m$-way $q$-shot, where $m$ determines the number of classes on each client and $q$ determines the number of data samples per class. We randomly set the classes possessed by each client and change the value of $m$ to adjust the heterogeneity degree of data distribution. 

\begin{figure*}[ht]
    \centering
		\subfloat[Proposed]{
            \label{scatter_FedContrast}
        \includegraphics[width=0.31\textwidth]{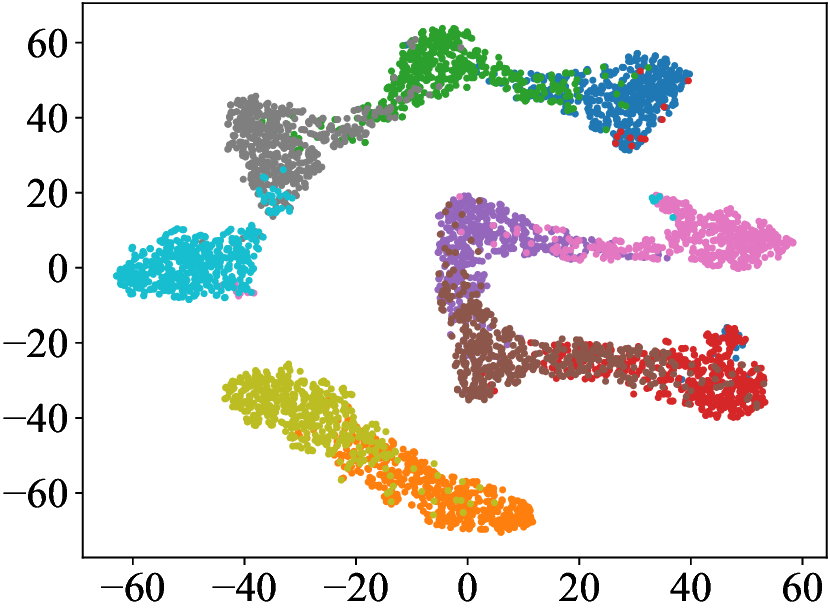}
		}
		\subfloat[FedProto]{
            \label{scatter_FedProto}
			\includegraphics[width=0.31\textwidth]{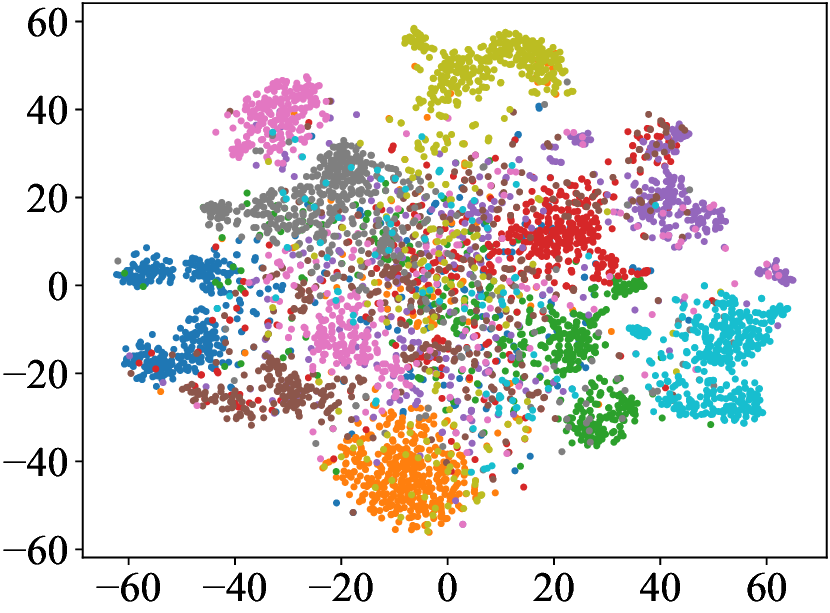}
		}
		\subfloat[FedAvg]{
            \label{scatter_FedAvg}
			\includegraphics[width=0.31\textwidth]{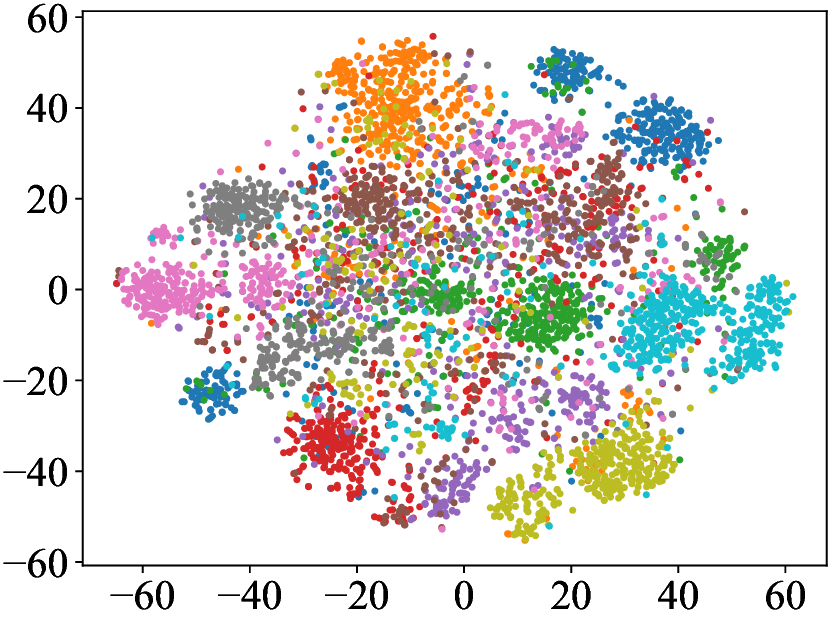}
		}
		\caption{The t-SNE visualization of semantic representations obtained by (a) proposed FedCL framework; (b) FedProto; (c) FedAvg.}
		\label{scatter}
\end{figure*}

Fig. \ref{performance1}\subref{test_acc} compares the learning performance of FedCL with other distributed learning benchmarks. With 20 clients, FedCL outperforms FedProto and FedAvg by $12.75\%$ and $38.55\%$, respectively. FedCL also remains stable with fewer clients, unlike FedProto, which degrades with only 5 clients due to insufficient training samples.
Fig. \ref{performance1}\subref{snr_acc} examines the impact of channel SNR on model performance across different schemes. At SNR $=9$ dB, FedCL achieves $25.09\%$ and $66.24\%$ higher accuracy than FedProto and FedAvg, respectively. Different from FedProto and FedAvg which suffer significant degradation at low SNR, the FedCL only degrades by approximately $10\%$, revealing that trainable global semantic centroids based on contrastive learning provide a better regularization for noised semantic features, which improve the robustness against the interference of wireless channel noise.

Fig. \ref{performance2}\subref{test_acc_m} illustrates FedCL's learning performance across varying $m$, reflecting the impact of data semantic heterogeneity on task-oriented SemCom. Note that the total amount of client data remains constant for fair comparison. A smaller $m$ indicates greater semantic heterogeneity, with FedCL demonstrating superior adaptability to such scenarios. The decrease in performance with increasing $m$ may result from insufficient training samples in each single category. 
Fig. \ref{performance2}\subref{m_acc} compares task performance across different schemes with varying $m$ when $K=5$ and SNR $=10$ dB. It demonstrates that the proposed scheme notably outperforms baselines, particularly under highly heterogeneous data distributions.

Fig. \ref{scatter} displays the distribution of noised semantic features from models trained under three different schemes after dimensionality reduction using t-SNE method with SNR $=5$ dB. The proposed FedCL demonstrates a unified semantic space, where the noised semantic features exhibit intra-semantic solidarity and inter-semantic discriminability. It preserves clearer semantic boundary against significant channel noise compared to benchmarks without trainable global semantic centroids, suggesting its superior separation capability based on the contrastive learning method.

\section{Conclusion}
In this letter, we proposed the FedCL framework for task-oriented communications, where personalized semantic encoders from multiple clients and a global semantic decoder at the BS were collaboratively learned. Unlike existing strategies that necessitated model structural consistency for aggregation, the proposed FedCL framework supported heterogeneous client-side semantic encoders. Additionally, we utilized contrastive learning to train the SCG for global semantic centroid generating, which regularizes heterogeneous local semantic features into discriminative global semantic space. The convergence analysis is also provided. Simulation results demonstrated that the proposed FedCL framework enhanced task performance and robustness compared to baseline schemes.

\bibliography{paper}
\bibliographystyle{IEEEtran}

\end{document}